\documentclass[twocolumn,prl,aps,notitlepage]{revtex4}
\usepackage{tikz}
\usepackage{amssymb}
\usepackage{psfrag}
\usepackage{textcomp}
\usepackage{pdfsync}
\usepackage{amsmath}
\usepackage{amsfonts}
\usepackage{graphicx,bm,epstopdf}
\usepackage{subfigure}
\usepackage{comment}
\usepackage{verbatim}
\usepackage{color}
\usepackage{upgreek}
\usepackage{tikz}
\usepackage{natbib,hyperref}
\usepackage{ifpdf}
\usepackage{float}

\begin{document}

\title{Spin-to-charge conversion in magnetic Weyl semimetals}
\author{Steven S.-L. Zhang$^{1}$}
\email{shulei.zhang@anl.gov}
\author{Anton A. Burkov$^{2}$}
\author{Ivar Martin$^{1}$}
\author{Olle G. Heinonen$^{1,3}$}
\affiliation{$^1$Materials Science Division, Argonne National Laboratory, Lemont,
Illinois 60439, USA\\
$^2$Department of Physics and Astronomy, University of Waterloo, Waterloo,
Ontario, N2L 3G1, Canada\\
$^3$Northwestern-Argonne Institute of Science and Engineering, Evanston,
Illinois 60208, USA}

\begin{abstract}
Weyl semimetals (WSMs) are a newly discovered class of quantum materials
which can host a number of exotic bulk transport properties, such as the
chiral magnetic effect, negative magnetoresistance, and the anomalous Hall
effect. In this work, we investigate theoretically the spin-to-charge
conversion in a bilayer consisting of a magnetic WSM and a normal metal (NM),
where a charge current can be induced in the WSM by an spin current injection at
the interface. We show that the induced charge current exhibits a peculiar
anisotropy: it vanishes along the magnetization orientation of the magnetic WSM, regardless of the direction of the injected spin. This anisotropy originates from the unique band structure of magnetic WSMs and
distinguishes the spin-to-charge conversion effect in WSM/NM structures from that
observed in other systems, such as heterostructures involving heavy metals
or topological insulators. The induced charge current depends strongly on
injected spin orientation, as well as on the position of the
Fermi level relative to the Weyl nodes and the separation between them.
These dependencies provide additional means to control and manipulate
spin-charge conversion in these topological materials.
\end{abstract}

\date{\today}
\maketitle

%\\
%$^4$Computation Institute, The University of Chicago, 5735 S Ellis Ave,
%Chicago, Illinois 60637 USA}

%\email{shulei.zhang@anl.gov}

\bigskip

Central to spintronics are the inter-conversion
between charge and spin currents, and the manipulation and detection of the
spin orientation of current-carrying itinerant electrons. Inter-conversion
can utilize the spin Hall effect (SHE)~\cite%
{DYAKONOV71PLA_spinHall,Hirsch99PRL_SHE,sZhang00PRL_SHE,Vignale10_SHE,Sinova15}
and the inverse spin Hall effect (ISHE) -- the Onsager reciprocal of the SHE
that originates in bulk spin-orbit interactions; the SHE
and ISHE are well-established phenomena which convert a charge current to a
spin current (SHE), and a spin current to a charge current propagating
perpendicularly to both the spin and flow directions of the injected spin
current (ISHE). The ISHE has been playing an important role in detecting
spin current generation in various heterostructures via transport
measurements~\cite{aHoffman13IEEE_SHE,Sinova15}.

Recently, several experimental and theoretical studies~\cite%
{Sanchez13NatCommn_SC-conversion,kShen14PRL_IEE,Saitoh14PRL_SC-TI,Lesne16NatMater_SC-2DEG,Sanchez16PRL_SC-conversion,sZhang&Fert16PRB_SC-TI,Chien18PRL_SC-Bi,Han18NPJ-QM_SC-review}
have investigated the inverse Edelstein effect (IEE) in an interfacial
two-dimensional electron gas (2DEG) with Rashba spin-splitting, or in a 2DEG at
the surface of a three-dimensional (3D) topological insulator. In the IEE a
spin accumulation in the 2DEG induces a charge current flowing
perpendicularly to the nonequilibrium spin orientation. Compared to the
ISHE that typically occurs in bulk systems, the spin-to-charge conversion
based on the IEE is arguably more efficient by taking advantage of the
remarkable spin-momentum locking arising from strong interfacial
spin-orbit coupling as well as of broken inversion symmetry at the interface
involving a heavy metal or topological insulator layer.

Weyl semimetals (WSMs), a newly discovered class of quantum materials, is
another rapidly evolving research field~\cite%
{Burkov16NatMater_SM-review,bhYan17ARCMP_WSM-review,Hasan17ARCMP_WSM-review,Armitage18RMP_SM}%
. This novel semimetal possesses distinct electronic properties,
such as the chiral anomaly~\cite%
{Adler69PR_chiral-anomaly,Bell&Jackiw69_Chiral-anomaly,NIELSEN83PLB_anomaly}
and Fermi arc surface states~\cite{xgWan11PRB_FermiArc}, that are
protected by the nontrivial topology of the band structure. WSMs studied to date
have broken inversion symmetry (but are time-reversal invariant) with at
least four, and often many more, Weyl nodes. The relatively large number of
Weyl nodes makes it difficult to clearly elucidate and control effects
related to the location of the Weyl nodes, and the lack of a magnetic order
parameter prevents direct coupling to magnetic fields. Recently, there has
been increasing interest in the pursuit of magnetic Weyl semimetals, which
can have only two Weyl
nodes present near the Fermi surface --- an ideal system to investigate transport
properties --- and also allow for direct coupling with external magnetic
field to control and manipulate electronic and transport properties.
Most efforts have been dedicated to seeking potential candidates of magnetic
Weyl semimetals~\cite%
{zjWang16PRL_magWSM,Kubler16EPL_magWSM,wjShi18PRB_magWSM,2019arXiv_lWang_magnWSM}
and examining their bulk transport properties. In contrast, little attention
has been paid to the coupled spin and charge degrees of freedom in
heterostructures composed of magnetic WSMs and other materials, which is of fundamental interest
and may be important for future applications of WSMs in spintronics.

\begin{figure}[htb]
\centering
\includegraphics[trim={0.0cm 0.cm 0.0cm 0.cm},clip=true,
width=1.0\linewidth]{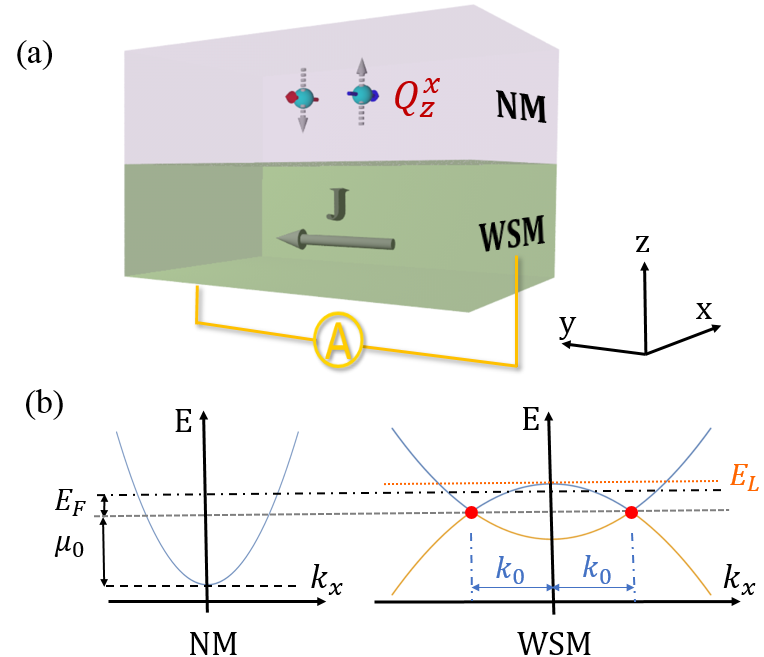}
\caption{Schematics of spin-to-charge conversion in a bilayer consisting of
a normal metal (NM) and a magnetic Weyl semimetal (WSM). Panel (a): A spin
current $Q_z^{\protect\alpha}$ flowing in the $z$-direction (perpendicular
to the layer plane) and with spin polarized in an arbitrary direction
(denoted by the superscript ``$\protect\alpha$") is generated in the NM
layer, which is subsequently converted to a charge current $\boldsymbol{J}$
in the WSM layer. Panel (b): sketches of band diagrams $E(\mathbf{k})$
with $k_y=k_z=0$ for the NM and the WSM with the dash-dotted line and the
dash line denoting respectively the Fermi energy $E_F$ and the conduction band
bottom of the NM with respect to the energy of the pair of
Weyl nodes (red dots) which are separated by $2k_0$; the orange dotted line denotes the Lifshitz transition energy $E_L$ above which two separate Fermi surfaces, enclosing the two  Weyl  nodes,  merge  into  a  single one. }
\label{fig1:schematics}
\end{figure}

In this work, we investigate theoretically the spin-to-charge conversion in
a bilayer consisting of a magnetic WSM with two Weyl nodes and a nonmagnetic
metal (NM). We use a scattering formalism to connect electronic states in
the NM and the magnetic WSM. Having obtained the scattering states, we use
semi-classical transport theory~\cite{morimoto16PRB_semi-cls-WSM} to demonstrate that
a charge current can be induced in the magnetic WSM by injecting a spin
current from the %We demonstrate that a distribution of charge current may be
%induced in the magnetic WSM layer by creating a spin accumulation at the
interface of the bilayer [see Fig.~\ref{fig1:schematics}(a)]. %, without specifying the spin
%injection mechanism,
Furthermore, we
show that the unique properties of magnetic WSMs allow for a control of the spin-to-charge
conversion that has no analog in conventional NM and magnetic bilayer
systems.

Let us commence with the following minimal model
Hamiltonian~\cite{sqSehn17,Armitage18RMP_SM} for the magnetic WSM layer filling the $z<0$ half space:
\begin{equation}
\mathcal{H}_{W}=\left[ m_{1}\left( k_{0}^{2}-k_{x}^{2}\right) +m_{0}\left(
k_{y}^{2}+k_{z}^{2}\right) \right] \sigma _{x}+v\left( k_{y}\sigma
_{y}+k_{z}\sigma _{z}\right) \,,  \label{eq:W_H}
\end{equation}%
where $\sigma _{i}$ ($i=x,y$ and $z$) are Pauli spin matrices, and $m_{0}$, $%
m_{1}$ and $v$ are generic materials parameters. Note that the two Weyl
nodes are located at $\mathbf{k=}\left( \pm k_{0},0,0\right) $ and that the
first term on the right-hand-side of Eq.~(\ref{eq:W_H}) breaks
time-reversal symmetry. The Weyl Hamiltonian has the eigenvalues
\begin{equation}
E_{\mathbf{k},s}=s\sqrt{\left[ m_{1}\left( k_{0}^{2}-k_{x}^{2}\right)
+m_{0}\left( k_{y}^{2}+k_{z}^{2}\right) \right] ^{2}+v^{2}\left(
k_{y}^{2}+k_{z}^{2}\right) }\,,  \label{eq:W_E}
\end{equation}%
where $s=\pm 1$ with $E_{\mathbf{k},+}$ and $E_{\mathbf{k},-}$
corresponding to the upper and lower energy bands that touch at the pair of
Weyl nodes. The $z\geq 0$ region is occupied by a NM described by the
Hamiltonian $\mathcal{H}_{N}=\frac{\boldsymbol{\hat{p}}^{2}}{2m_{e}}-\mu
_{0} $, where $m_{e}$ is the effective mass and $\mu _{0}$ denotes the
deviation of the conduction band bottom of the NM layer from the energy of the two Weyl nodes, as sketched in Fig.~\ref%
{fig1:schematics}(b). Note that we will only consider $\mu _{0}>0$
so that there are available scattering states in the NM when the two Weyl
nodes are in the close vicinity of the Fermi energy, which are the
circumstances under which most of the interesting transport phenomena in
WSMs emerge~\cite%
{yRan11PRB_Hall-WSM,Burkov14PRL_AHE-WSM,Burkov15PRB_MR-WSM,sqShen16NJP_MR-WSM}%
.

In the NM, by choosing the $z$-axis as the spin quantization axis, the full
scattering wave function for a free electron with a given energy $E$ and a
spin pointing in an arbitrary direction $\boldsymbol{n}\mathbf{=}\left( \sin
\theta \cos \phi ,\sin \theta \sin \phi ,\cos \theta \right) $ can be
written as a linear combination of spin-up and spin-down components, i.e.,
\begin{align}
\varphi _{N}(\mathbf{k},\mathbf{r})& =\left[ \cos \frac{\theta }{2}e^{-i\phi /2}\binom{1%
}{0}\left( e^{-ik_{z}z}+R_{\uparrow }e^{ik_{z}z}\right) \right.  \notag \\
& \left. +\sin \frac{\theta }{2}e^{i\phi /2}\binom{0}{1}\left(
e^{-ik_{z}z}+R_{\downarrow }e^{ik_{z}z}\right) \right] e^{i\mathbf{k}%
_{\parallel }\cdot \mathbf{r}}\,,
\end{align}%
where $k_{z}\equiv \sqrt{\frac{2m_{e}E}{\hbar ^{2}}-\mathbf{k}%
_{\parallel }^{2}}$ with $\sigma =\uparrow \left( \downarrow \right) $ and $%
\mathbf{k}_{\parallel }\left[ \equiv \left( k_{x},k_{y}\right) \right] $
the in-plane component of the wavevector, $R_{\sigma }$ are the reflection
amplitudes, and we have assumed translational invariance in the $x$-$y$
plane.

The wave function for an electron transmitted into the magnetic WSM can be
expressed as
\begin{equation}
\varphi _{W}\left(\mathbf{k},\mathbf{r}\right) =\left( T_{+}\chi
_{+}e^{ik_{z,+}z}+T_{-}\chi _{-}e^{ik_{z,-}z}\right) e^{i\mathbf{k}%
_{\parallel }\cdot \mathbf{r}}
\end{equation}%
where $T_{\pm }$ are the transmission amplitudes, $\chi _{+}$ and $\chi _{-}$
are two spinors given by $\chi _{\pm }=\frac{1}{\sqrt{N_{\pm }}}\binom{%
a_{\pm }}{b_{\pm }}$ with $a_{\pm }=m_{1}\left( k_{x}^{2}-k_{0}^{2}\right)
+m_{0}\left( k_{y}^{2}+k_{z,\pm }^{2}\right) -ivk_{y}$, $b_{\pm
}=E-vk_{z,\pm }$ and $N_{\pm }$ the normalization coefficients satisfying $%
\left\vert a_{\pm }\right\vert ^{2}+\left\vert b_{\pm }\right\vert
^{2}=N_{\pm }^{2}$. The $z$-components of the wavevectors are given by
\begin{eqnarray}
k_{z,\pm }^{2} &=&-\frac{1}{m_{0}^{2}}\left\{ m_{0}m_{1}\left(
k_{0}^{2}-k_{x}^{2}\right) +m_{0}^{2}k_{y}^{2}+\frac{v^{2}}{2}\right. \,
\notag \\
&&\left. \mp \sqrt{m_{0}m_{1}\left( k_{x}^{2}-k_{0}^{2}\right)
v^{2}+m_{0}^{2}E^{2}+\frac{v^{4}}{4}}\right\}
\end{eqnarray}%
with the signs of $k_{z,\pm }$ so selected that the transmitted waves either
propagate freely or decay in the WSM ($z<0$).

The reflection and transmission amplitudes $R_{\uparrow \left( \downarrow
\right) }$ and $T_{\pm }$ can be determined by proper boundary conditions.
Here, we assume that both the scattering wave function and the $z$-component
of the current density are continuous at the interface $z=0$:
\begin{subequations}
\begin{equation}
\varphi _{N}(0^{+})=\varphi _{W}\left(0^{-}\right)
\end{equation}%
\begin{equation}
\hat{v}%
_{N,z}\varphi _{N}(0^{+})=\hat{v}_{W,z}\varphi _{W}(0^{-})
\end{equation}%
\end{subequations}
where the velocity operators are given by $\boldsymbol{\hat{v}}_{N}=\frac{%
\partial \mathcal{H}_{N}}{\hbar\partial \mathbf{k}}$ and $\boldsymbol{\hat{v}}%
_{W}=\frac{\partial \mathcal{H}_{W}}{\hbar\partial \mathbf{k}}$ for the NM
and WSM, respectively, and we have eliminated a common phase factor of $e^{i\mathbf{k}_{\parallel}\cdot\mathbf{r}}$. We shall present the full expressions of the scattering amplitudes in the Supplemental Materials as they are a bit lengthy and are not very instructive.

\begin{figure*}[t!]
\centering
\includegraphics[trim={0.8cm 7.5cm 0.7cm 6cm},clip=true,
width=1.0\linewidth]{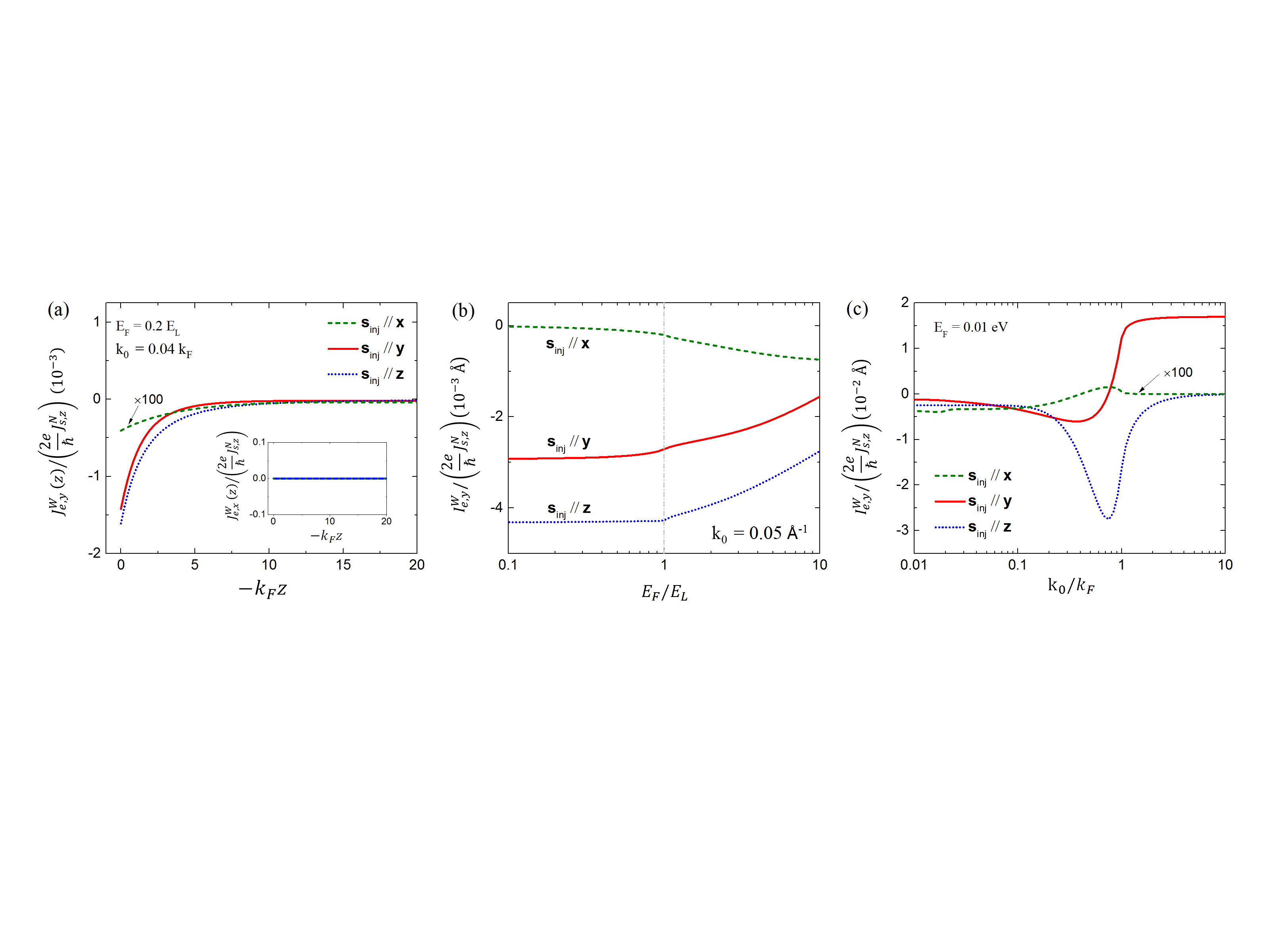}
\caption{Characterization of the induced charge current in the magnetic WSM
layer: (a) Current density $\boldsymbol{J}_{e,\parallel}^W$ as a function of
the spatial coordinate $z$, the spatially integrated current $\boldsymbol{I}%
_{e,\parallel}^W \left(\equiv \protect\int_{-\infty}^0 \mathit{d}z%
\boldsymbol{J}_{e,\parallel}^W \right)$ as functions of (b) Fermi energy $%
E_F $ and (c) the separation of the two Weyl nodes $k_0$ for an injected
spin $\boldsymbol{s}_{inj}$ along $x$, $y$ and $z$ axes respectively. The
Lifshitz transition energy and the Fermi wavevector are given by $E_L=m_0
k_0^2$ and $k_F=\protect\sqrt{2m_e(E_F+\protect\mu_0)/\hbar^2}$,
respectively. Other parameters used in the numerical calculation: $\protect%
\mu_0=5.0$ eV, $m_e=9 \times 10^{-31}$ kg, $m_0=-m_1=20$ eV$\cdot \mathring{A%
}^2$ and $v=2$ eV$\cdot \mathring{A}$. }
\label{fig2:J_W}
\end{figure*}

Thus far, we have solved the problem of a single electron scattering at the
interface of the magnetic-WSM/NM bilayer. In order to calculate the current
density induced in the magnetic WSM layer due to the spin current injection from the NM layer, we need to find the electron distributions in each layer.
At the NM side of the interface (i.e., $z=0^{+}$), the distribution function
can be described by a $2\times 2$ matrix in spin space~\cite%
{mStiles02PRB_STT,jwZhang04PRL,slzhang15PRB_AMR,slZhang16PRL}, i.e.,
\begin{equation}
\hat{f}_{N}=f_{0,N}\left( \mathbf{k}\right) \hat{I}+\hat{g}_{N}\left(
\mathbf{k}\right)
\end{equation}%
where $f_{0,N}\left( \mathbf{k}\right) \hat{I}$ is the equilibrium part
of the distribution function with $f_{0,N}$ the Fermi-Dirac function and $%
\hat{I}$ the $2\times 2$ identity matrix, and the nonequilibrium component
of the distribution function $\hat{g}_{N}\left( \mathbf{k}\right) $ that gives rise to the spin current can be described by
\begin{equation}
\hat{g}_{N}\left( \mathbf{k}\right) =-e\tau v_{z}{\hat{\mathcal{E}}}_{z}%
\frac{\partial f_{0}}{\partial E_{\mathbf{k}}}
\end{equation}%
where ${\hat{\mathcal{E}}}_{z}=E_{z}\boldsymbol{\sigma \cdot s}_{inj}$ (with
$\boldsymbol{\sigma }$ denoting the Pauli spin matrices) is a spin-dependent
electric field pointing in opposite directions for electrons with opposite
spin directions which drives a spin current~\footnote{The spin current can be realized experimentally in a few different ways, for example through the SHE~\cite{aHoffman13IEEE_SHE} or through spin pumping~\cite{Tserkovnyak05RMP}; the manner through which the spin current is generated is not important for our purposes so we write the spin current in terms of an effective spin-dependent electric field.}, $\boldsymbol{s}_{inj}$ is a
unit vector denoting the direction of the spin component of the spin
current. At temperatures well below the Fermi temperature of the NM, it is a
good approximation to assume $\frac{\partial f_{0,N}}{\partial E_{%
\mathbf{k}}}\simeq -\delta \left( E_{\mathbf{k}}-\mu
_{0}-E_{F}\right) $ where $E_{F}$ is the Fermi energy relative to the energy of the two Weyl nodes as shown schematically in Fig.~\ref%
{fig1:schematics}(b). Formally, the spin current density is given by $%
Q_{z}^{\alpha }=\frac{\hbar }{4}Tr_{\sigma }\int \frac{d^{3}\mathbf{k}}{%
\left( 2\pi \right) ^{3}}\sigma ^{\alpha }v_{z}\hat{f}_{N}$. Explicitly, $%
Q_{z}^{b}=J_{s,z}^{N}s_{inj}^{b}$ where the magnitude of the spin current
density for a given spin direction can be characterized by $%
J_{s,z}^{N}\equiv \frac{\hbar }{2e}\sigma _{D}E_{z}$ with $\sigma _{D}=\frac{%
\tau e^{2}k_{F}^{3}}{3\pi ^{2}m_{e}}$ the Drude conductivity.

The nonequilibrium distribution function for the transmitted electrons in the magnetic WSM is
determined by the transmission amplitudes and the nonequilibrium electron
distribution $\hat{g}_{N}\left( \mathbf{k}\right) $ at $z=0^{+}$~\cite%
{camley89PRL,mStiles02PRB_STT} via
\begin{equation}
\hat{g}_{W}^{<}\left( \mathbf{k},z\right) =\hat{T}^{\dag }\left( \mathbf{k},z%
\right) \hat{g}_{N}^{<}\left( \mathbf{k}\right) \hat{T}\left( \mathbf{k},z%
\right)
\end{equation}%
where $\hat{T}\left( \mathbf{k},z\right) $ is a $2\times 2$ transmission
matrix satisfying $\varphi_W(\mathbf{k},z)=\hat{T}(\mathbf{k},z)\varphi_{N,i}(\mathbf{k},0^{+})$ (the formula for $\hat{T}\left( \mathbf{k},z\right)$ is not very informative and thus we will present
it together with the derivation of the scattering amplitudes in the Supplemental Materials), the superscript ``$<$" denotes electrons moving
in the negative $z-$direction (i.e., $v_z<0$). Note that, to the leading order, electrons in the WSM moving towards the interface are assumed to entirely come from the equilibrium distribution, i.e., $\hat{f}_{W}^{<}\simeq \hat{f}_{0,W}^{<}$ and $\hat{g}_{W}^{<}\simeq 0$.

Having obtained the nonequilibrium distribution $\hat{g}_{W}\left( \mathbf{%
k},z\right) $, the in-plane charge current density induced in the magnetic
WSM layer can be computed via
\begin{equation}
\boldsymbol{J}_{e,\parallel }^{W}(z)=-\frac{e}{2}\int \frac{d^{3}\mathbf{k}%
}{\left( 2\pi \right) ^{3}}Tr_{\sigma }\left( \hat{g}_{W}\boldsymbol{\hat{v}}%
_{W,\parallel }+\mathit{h.c.}\right) \,,  \label{Eq: J_W}
\end{equation}%
where $\mathit{h.c.}$ denotes Hermitian conjugate, $\hat{g}_{W}=\hat{g}_{W}^{>}+\hat{g}_{W}^{<}$ and the trace operation is
carried out in the spin space. Note that in deriving Eq.~(\ref{Eq: J_W}) we have taken into account the fact that the equilibrium distribution of electrons in the WSM, i.e., $\hat{f}_{0,W}$, does not contribute to the current.

Before seeking the numerical solutions of
the induced charge current in the magnetic WSM layer, a remarkable property
of the spin-to-charge conversion can be illuminated by a simple symmetry
analysis of Eq.~(\ref{Eq: J_W}): regardless of the orientation of the
injected spins, no current will be induced in the direction parallel to
the line connecting the pair of Weyl nodes in momentum space, i.e., $%
J_{e,x}^W=0$. This is simply because the $x$-component of the electron
velocity operator (i.e., $\hat{v}_{W,x}=\frac{\partial \mathcal{H}_{W}}{%
\partial k_{x}}$) is an odd function of $k_{x}$ whereas the nonequilibrium
distribution function $\hat{g}_{W}$ is an even function of $k_{x}$. Therefore,
the corresponding $x$-component of the current density must vanish everywhere in the magnetic WSM layer as it is the integral of the product of these two over $\mathbf{k}$%
-space. Such an
anisotropic spin-to-charge conversion stems from the inherent property of
magnetic WSMs -- the anisotropy in the band structure in the first Brillouin
zone between the directions perpendicular and parallel to the separation
between the two Weyl nodes in $\mathbf{k}$-space. The numerical solution
of $J_{e,x}^{W}$ indeed confirms that it vanishes everywhere in the magnetic
WSM layer, regardless of the direction of the injected spin $\boldsymbol{s}%
_{inj}$, the position of the Fermi level $E_{F}$ as well as the separation
between the Weyl nodes [see the inset of Fig.~\ref{fig2:J_W}(a)].

In contrast to the robust suppression of $J_{e,x}^{W}$, the behavior of the
current induced in the $y$-direction (perpendicular to the separation
between the two Weyl nodes and perpendicular to the WSM$\mid$NM interface) is much richer.
Figure~\ref{fig2:J_W}(a) shows
the spatial variation of the $J_{e,y}^{W}(z)$ for the injected spin along $x
$, $y$ and $z$ directions, respectively. We find that while the magnitude of $%
J_{e,y}^{W}$ depends on the orientation of the spin injection, it generally
decays rapidly over one Fermi wavelength $\lambda _{F}\left( =\frac{2\pi }{%
k_{F}}\right) $ away from the WSM$\mid$NM interface, indicating a dominant contribution of the evanescent surface
states to the spin-to-charge conversion in the magnetic WSM.

In Fig.~\ref{fig2:J_W}(b), we show the total induced current $%
I_{e,y}^{W}\left( \equiv \int_{-\infty }^{0}\mathit{d}zJ_{e,y}^{W}\right) $
as a function of the Fermi level $E_{F}$. We note the existence of a
Lifshitz transition energy level $E_{L}$ at which two separate Fermi
surfaces, enclosing the two Weyl nodes, merge into a single Fermi surface
[as shown schematically in Fig.~\ref{fig1:schematics}(b)]. We find that the
total current $I_{e,y}^{W}$ is insensitive to the variation of the
Fermi level as long as $E_{F}$ is below $E_{L}$, and the onset of noticeable
changes of $I_{e,y}^{W}$ occur at $E_{L}$ due to a significant change of
density of states at the Fermi level when it crosses the Lifshitz energy. In Fig.~\ref{fig2:J_W}(c), we
show the dependence of the total induced current $I_{e,y}^{W}$ on the
separation between the two Weyl nodes $2k_{0}$. Extraordinary variations of $%
I_{e,y}^{W}$ take place when $k_{0}$ approaches the Fermi wavevector $k_{F}$%
, as the projected Fermi contour of the NM in the $x$-$y$ plane switches
between one that encloses the two Weyl nodes and one that does not, which
drastically alters the scattering phase space.
Lastly, we provide an order-of-magnitude estimation of the effect in EuCd$_2$As$_2$ -- a magnetic Weyl predicted recently~\cite{2019arXiv_lWang_magnWSM} that contains a single pair of Weyl nodes. By choosing the following parameters for EuCd$_2$As$_2$~\cite{2019arXiv_lWang_magnWSM}: $m_0=1.6$ eV$\cdot \mathring{A%
}^2$, $m_1=54.5$ eV$\cdot \mathring{A%
}^2$, $v=2.7$ eV$\cdot \mathring{A%
}$, $k_0=0.008$ $\mathring{A%
}^{-1}$, and $E_F=0.01$ eV,
we obtain a spin-to-charge conversion efficiency of $\vartheta \simeq 0.2\%$ for spin injected along the $y$-direction where $\vartheta \equiv J_{e,y}^W(0^{-})/(\frac{2e}{\hbar}J_{s,z}^N)$, which is about an order of magnitude smaller than the spin Hall angle in Pt~\cite{aHoffman13IEEE_SHE,Sinova15}.

As a final point, it is interesting to compare the spin-to-charge conversion in magnetic WSMs
with that in other systems (such as heavy metals, Rashba 2DEG, topological
insulator surfaces etc.) due to the ISHE or IEE. When a spin current is
injected in heavy metals (such as Pt or Ta) from a NM, a charge current will
be generated in the direction perpendicular to both the spin direction and
and the flow direction of the injected spin current due to the ISHE; formally,
the process can be described by $J_{e,i}=\epsilon _{ijk}\vartheta
_{0}Q_{j}^{k} $ where $\epsilon _{ijk}$ is the antisymmetric Levi-Civita
tensor ($i,j,k=x$,$y$, or $z$), $Q_{j}^{k}$ represents the injected spin
current flowing along the $j$ direction with spin pointing in the $k$
direction, and $\vartheta _{0}$ is a dimensionless material parameter known
as the spin Hall angle which measures the efficiency of the spin-charge
conversion. A transverse charge current can also be generated, based on the
inverse Edelstein effect, by injecting a spin current perpendicularly to the
surface of a topological insulator or to an interface with strong Rashba
spin-orbit coupling and using the spin-charge locking in these systems that
fixes the spins of the carriers perpendicularly to their momenta. Note that
the IEE has the same symmetry as the ISHE and hence can be described by the
same formula that we used for the ISHE -- the only difference is that it is
a conversion of a 3D spin current to a 2D charge current and hence $%
\vartheta _{0}$ in the linear response relation has the dimension of length.
For both IEE and ISHE, a charge current may in principle be induced in any
arbitrary direction with properly chosen spin injection direction, i.e., $%
\vartheta _{0}$ is \textit{isotropic}~\cite{slZhang14EPL}.

The spin-to-charge conversion in magnetic WSMs, however, is rather \textit{%
anisotropic} emanating from the anisotropy in their unique band structures
-- the appearance of a pair of Weyl nodes in $\mathbf{k}$-space; as we
have shown above, no charge current can be induced in the direction along
the line connecting the two Weyl nodes (i.e., $\mathbf{\hat{k}}_{0}$),
regardless of the orientation of the injected spins. Note that for a magnetic WSM with a single pair of Weyl nodes, the magnetization is in the same direction as $\hat{\mathbf{k}}_0$~\cite{2019arXiv_lWang_magnWSM}. In general, there will be an odd number of pairs
in a magnetic Weyl semimetal, in which case the total current density, being the sum of contributions from different pairs (if the pairs of Weyl nodes are well separated in the reciprocal space), vanishes along the magnetization direction, i.e.,
\begin{equation}
\mathbf{m} \cdot \boldsymbol{J}_{e,\parallel }^{W}=0 \,,
\end{equation}
where $\mathbf{m}$ is a unit vector denoting the magnetization direction of the magnetic WSM.

A charge current, however, can be induced in the direction perpendicular to the magnetization direction, and
the induced current is rather sensitive to the direction of the injected spin $\mathbf{s}_{inj}$, which is experimentally controllable. In addition, we have shown that the spin-to-charge conversion in magnetic
WSM relies on the separation between two Weyl nodes and the position of the
Fermi surface relative to them, which provides additional means to
manipulate and control the effect. These remarkable features make the
spin-to-charge conversion in magnetic WSMs distinctly different from that
previously studied in heterostructures involving heavy metals or topological
insulators, and are potentially very useful in spintronic applications.

Work by S. Z., A. B. and O.H. was supported by Center for Advancement of
Topological Semimetals, an Energy Frontier Research Center funded by the
U.S. Department of Energy Office of Science, Office of Basic Energy
Sciences, through the Ames Laboratory under its Contract No.
DE-AC02-07CH11358; work by I. M. was supported by the U.S. DOE, Office of
Science, Basic Energy Science Division of Materials Sciences and Engineering.

\appendix

\section{Appendix:~Derivation of the transmission matrix in spin space}

Let us first consider the scattering problem for free electrons in
a bilayer consisting of a normal metal (NM) layer and a magnetic Weyl
semimetal (WSM) layer, and with the electrons incident on the interface from the NM. For the Weyl fermions in the magnetic WSM layer, we
use the following low-energy effective Hamiltonian [same as Eq. (1) in the
main text]
\begin{equation}
\mathcal{H}_{W}=\left[ m_{1}\left( k_{0}^{2}-k_{x}^{2}\right) +m_{0}\left(
k_{y}^{2}+k_{z}^{2}\right) \right] \sigma _{x}+v\left( k_{y}\sigma
_{y}+k_{z}\sigma _{z}\right) \,,  \tag{S1}
\end{equation}%
where $\sigma _{i}$ ($i=x,y,z$) are the Pauli spin matrices, and for the NM
layer we adopt the following simple free electron model Hamiltonian
\begin{equation}
\mathcal{H}_{N}=\frac{\hbar ^{2}k^{2}}{2m_{e}}-\mu _{0} \,,   \tag{S2}
\end{equation}%
where $\mu _{0}$ is a constant shift of the chemical potential.

By choosing the $z$-axis as the spin quantization axis, the wave function of an electron in the NM layer incident on the interface and with its spin pointing in an arbitrary
direction $\boldsymbol{n}\mathbf{=}\left( \sin \theta \cos \phi ,\sin \theta
\sin \phi ,\cos \theta \right) $ can be written as
\begin{equation}
\varphi _{N,i}(\mathbf{r})=\left[ \cos \frac{\theta }{2}e^{-i\phi /2}\binom{1%
}{0}+\sin \frac{\theta }{2}e^{i\phi /2}\binom{0}{1}\right] e^{-ik_{z}z}e^{i%
\boldsymbol{k}_{\parallel }\cdot \mathbf{r}}\,,  \tag{S3}  \label{Eq: phi_Ni}
\end{equation}%
and the corresponding reflected wave can be expressed as
\begin{equation}
\varphi _{N,r}(\mathbf{r})=\left[ R_{\uparrow }\binom{1}{0}+R_{\downarrow }%
\binom{0}{1}\right] e^{ik_{z}z}e^{i\boldsymbol{k}_{\parallel }\cdot \mathbf{r%
}}\,,  \tag{S4}
\end{equation}%
where $k_{z}\equiv \sqrt{\frac{2m_{e}E}{\hbar ^{2}}-\boldsymbol{k}%
_{\parallel }^{2}}$ with $%
\boldsymbol{k}_{\parallel }\left[ \equiv \left( k_{x},k_{y}\right) \right] $
the in-plane component of the wavevector, $R_{\uparrow(\downarrow) }$ are the reflection
amplitudes, and we have assumed translational invariance in the $x$-$y$
plane. It follows that the full scattering wave function in the NM is a
superposition of the incident and the reflected waves, i.e.,
\begin{equation}
\varphi _{N}(\mathbf{r})=\varphi _{N,i}(\mathbf{r})+\varphi _{N,r}(\mathbf{r}%
)\,.  \tag{S5}
\end{equation}

The wave function for a transmitted electron in the magnetic WSM can be
expressed as
\begin{equation}
\varphi _{W}\left( \mathbf{r}\right) =\left( T_{+}\chi
_{+}e^{ik_{z,+}z}+T_{-}\chi _{-}e^{ik_{z,-}z}\right) e^{i\boldsymbol{k}%
_{\parallel }\cdot \mathbf{r}} \,, \tag{S6}  \label{Eq:phi_W}
\end{equation}%
where $T_{\pm }$ are the transmission amplitudes, $\chi _{+}$ and $\chi _{-}$
are two spinors given by $\chi _{\pm }=\frac{1}{\sqrt{N_{\pm }}}\binom{%
a_{\pm }}{b_{\pm }}$ with $a_{\pm }=m_{1}\left( k_{x}^{2}-k_{0}^{2}\right)
+m_{0}\left( k_{y}^{2}+k_{z,\pm }^{2}\right) -ivk_{y}$, $b_{\pm
}=E-vk_{z,\pm }$ and $N_{\pm }$ the normalization coefficients satisfying $%
\left\vert a_{\pm }\right\vert ^{2}+\left\vert b_{\pm }\right\vert
^{2}=N_{\pm }^{2}$. The $z$-components of the wavevectors are given by
\begin{widetext}
\begin{equation}
k_{z,\pm }^{2}=-\frac{1}{m_{0}^{2}}\left[ m_{0}m_{1}\left(
k_{0}^{2}-k_{x}^{2}\right) +m_{0}^{2}k_{y}^{2}+\frac{v^{2}}{2}\mp \sqrt{%
m_{0}m_{1}\left( k_{x}^{2}-k_{0}^{2}\right) v^{2}+m_{0}^{2}E^{2}+\frac{v^{4}%
}{4}}\right] \,  \tag{S7}
\end{equation}%
\end{widetext}
with the signs of $k_{z,\pm }$ so selected that the transmitted waves either
propagate freely or decay in the WSM ($z<0$).

The reflection and transmission amplitudes $R_{\uparrow \left( \downarrow
\right) }$ and $T_{\pm }$ can be determined by specifying the boundary conditions.
Here, we assume that both the scattering wave function and the $z$-component
of the current density are continuous at the interface $z=0$:
\begin{equation}
\varphi _{N}(0^{+})=\varphi _{W}\left( 0^{-}\right) \text{ and }\hat{v}%
_{N,z}\varphi _{N}(0^{+})=\hat{v}_{W,z}\varphi _{W}(0^{-}) \,,  \tag{S8}
\label{Eq: bc's}
\end{equation}%
where the velocity operators are given by $\boldsymbol{\hat{v}}_{N}=\frac{%
\partial \mathcal{H}_{N}}{\hbar\partial \boldsymbol{k}}$ and $\boldsymbol{\hat{v}}%
_{W}=\frac{\partial \mathcal{H}_{W}}{\hbar\partial \boldsymbol{k}}$ for the NM
and WSM, respectively, where we have eliminated the common factor $e^{i\boldsymbol{k}_{\parallel}\cdot \mathbf{r}}$ on both sides of each equation.

By placing Eqs.~(\ref{Eq: phi_Ni}) - (\ref{Eq:phi_W}) in Eq.~(\ref{Eq: bc's}%
), one can derive the following scattering amplitudes
\begin{subequations}
\label{Eq:R_up-dn}
\begin{equation}
R_{\uparrow }=a_{+}T_{+}+a_{-}T_{-}-\cos \frac{\theta }{2}e^{-i\phi /2}
\tag{S9a}  \label{Eq:R_up}
\end{equation}%
\begin{equation}
R_{\downarrow }=b_{+}T_{+}+b_{-}T_{-}-\sin \frac{\theta }{2}e^{i\phi /2}
\tag{S9b}\,,   \label{Eq:R_dn}
\end{equation}%
\end{subequations}
where
\begin{subequations}
\label{T_pm}
\begin{equation}
T_{+}=\frac{2\left( B_{-}\cos \frac{\theta }{2}e^{-i\phi /2}-A_{-}\sin \frac{%
\theta }{2}e^{i\phi /2}\right) }{A_{+}B_{-}-A_{-}B_{+}}  \tag{S10a}
\label{Eq:T+}
\end{equation}%
\begin{equation}
T_{-}=\frac{2\left( A_{+}\sin \frac{\theta }{2}e^{i\phi /2}-B_{+}\cos \frac{%
\theta }{2}e^{-i\phi /2}\right) }{A_{+}B_{-}-A_{-}B_{+}}  \tag{S10b}
\label{Eq:T_}
\end{equation}%
\end{subequations}
with
\begin{subequations}
\begin{equation}
A_{s}=\left( 1-\frac{m_{e}va}{\hbar ^{2}k_{N,z}}\right) a_{s}-\frac{%
2m_{0}a^{2}m_{e}k_{z,s}}{\hbar ^{2}k_{N,z}}b_{s}  \tag{S11a}
\end{equation}%
\begin{equation}
B_{s}=\left( 1+\frac{m_{e}va}{\hbar ^{2}k_{N,z}}\right) b_{s}-\frac{%
2m_{0}a^{2}m_{e}k_{z,s}}{\hbar ^{2}k_{N,z}}a_{s}  \tag{S11b}
\end{equation}%
\end{subequations}
and $s$ denoting $+$ or $-$. We note that $R_{\uparrow \left( \downarrow
\right) }$ and $T_{\pm }$ contain only quadratic terms of $k_{x}^{2}$ and $%
k_{y}^{2}$ -- a property that is useful in determining the presence of the
in-plane charge current in a given direction.

Next, we determine the transmission matrix by rewriting the transmitted
state in the form of $\varphi _{W}\left(z\right) =\hat{T}%
_{NW}\varphi _{N,i}\left( 0^{+}\right) $. Inserting the expressions of $%
T_{\pm }$ [Eqs.~(\ref{Eq:T+}) and (\ref{Eq:T_})] in Eq.~(\ref{Eq:phi_W}),
one can rewrite $\varphi _{W}\left( z\right) $ as
\begin{widetext}
\begin{equation}
\varphi _{W}\left(z\right) =\frac{2}{A_{+}B_{-}-A_{-}B_{+}}\left(
\begin{array}{cc}
a_{+}B_{-}e^{ik_{z,+}z}-a_{-}B_{+}e^{ik_{z,-}z} &
a_{-}A_{+}e^{ik_{z,-}z}-a_{+}A_{-}e^{ik_{z,+}z} \\
b_{+}B_{-}e^{ik_{z,+}z}-b_{-}B_{+}e^{ik_{z,-}z} &
b_{-}A_{+}e^{ik_{z,-}z}-b_{+}A_{-}e^{ik_{z,+}z}%
\end{array}%
\right) \binom{\cos \frac{\theta }{2}e^{-i\phi /2}}{\sin \frac{\theta }{2}%
e^{i\phi /2}} \,. \tag{S12}
\end{equation}%
\end{widetext}
It follows that the transmission matrix can be expressed as
\begin{equation}
\hat{T}_{NW}=\frac{2}{A_{+}B_{-}-A_{-}B_{+}}\sum\limits_{s=\pm
}se^{ik_{z,s}z}\binom{a_{s}}{b_{s}}\otimes \left(
\begin{array}{cc}
B_{-s} & -A_{-s}%
\end{array}%
\right) \,,  \tag{S13}
\end{equation}%
where $\otimes $ denotes a direct product.

\bigskip

\bigskip

\bigskip

\bibliographystyle{apsrev4-1}
\bibliography{20190321_IEE-WSM}

\appendix

%\begin{figure}[h]
%\includegraphics[trim={0.8cm 0.5cm 3cm 2.2cm},clip=true,
%width=0.9\columnwidth]{20180108_rhoTH-v-lambda_1skyrm-bbl.pdf} \hspace*{\fill%
%}
%\caption{Topological Hall resistivity $\protect\rho _{TH}$ generated by a
%single skyrmion bubble as a function of spin diffusion length $\protect%
%\lambda _{sd}$ in a thin film of width $w=6r_{sk}$ for several different $p_{%
%\protect\tau }$; the insets show the corresponding spatial profile of the
%emergent magnetic field. We have defined $\protect\rho _{TH}^{\left(
%0\right) }=\left( p_{\protect\tau }+p_{\protect\sigma }\right) R_{H}\protect%
%\psi _{0}/S$ and used $a_{DW}=0.2r_{sk}$ and $J_{ex}=0.2\protect\epsilon %
%_{F} $.}
%\label{Fig:rho-1Skrm-bbl}
%\end{figure}

\end{document}